\documentclass[aps,prx,superscriptaddress,nofootinbib,twocolumn]{revtex4-1}

\usepackage{mathrsfs}
\usepackage{amsfonts}
\usepackage{amssymb}
\usepackage{amsmath}
\usepackage{graphicx}
\usepackage[usenames,dvipsnames]{color}
\usepackage[colorlinks=true,citecolor=blue,linkcolor=magenta]{hyperref}
\usepackage{ulem}
\usepackage{lmodern}

\newcommand{\figpath}{./}

\newcommand{\abs}[1]{| #1 |}

\newcommand{\mean}[1]{\langle #1 \rangle}

\newcommand{\SIG}[2]{\sigma^{#1}_{#2}}
\newcommand{\sss}{\mathbf{s}}
\newcommand{\one}{\mathbf{1}}

\begin{document}

\title{Learning time-dependent noise to reduce logical errors:
\\Real time error rate estimation in quantum error correction}

\author{Ming-Xia Huo}

\affiliation{Beijing Computational Science Research Center, Beijing 100193, China}

\author{Ying Li}

\affiliation{Graduate School of China Academy of Engineering Physics, Beijing 100193, China}

\affiliation{Department of Materials, University of Oxford, Parks Road, Oxford OX1 3PH, United Kingdom}

\date{\today}

\begin{abstract}
Quantum error correction is important to quantum information processing, which allows us to reliably process information encoded in quantum error correction codes. Efficient quantum error correction benefits from the knowledge of error rates. We propose a protocol for monitoring error rates in real time without interrupting the quantum error correction. Any adaptation of the quantum error correction code or its implementation circuit is not required. The protocol can be directly applied to the most advanced quantum error correction techniques, e.g.~surface code. A Gaussian processes algorithm is used to estimate and predict error rates based on error correction data in the past. We find that using these estimated error rates, the probability of error correction failures can be significantly reduced by a factor increasing with the code distance.
\end{abstract}

\maketitle

\section{Introduction}

Quantum error correction~\cite{Nielsen2010} is crucial to long-time quantum memory and large-scale quantum computation. Without quantum error correction, usually the performance of a physical quantum system is not good enough for exploiting the full power of quantum information processing. The state-of-the-art life time of quantum states varies from 100 microseconds for superconducting qubits~\cite{Devoret2013} to hours for impurity spins in silicon~\cite{Saeedi2013}, which are not comparable with most daily used classical memory devices. In the leading experiments a thousand entangling gates can be performed on two ion qubits with only one error on average~\cite{Lucas, Wineland}, however implementing Shor's algorithm on a meaningful scale may require trillions of gates~\cite{Fowler2012, Joe2016}. Given the limited coherence time and gate fidelity of the physical system, quantum error correction can reduce errors on the logical information to an arbitrarily low level, so that quantum memory with extendible life time and scalable quantum computation becomes possible.

Efficient quantum error correction benefits from the knowledge of error rates~\cite{Poulin2006, DuclosCianci2010, DuclosCianci2010arXiv, DuclosCianci2014, Stace2010, Wang2011}. In quantum error correction, the quantum information is encoded in a subspace, i.e.~the logical subspace. Errors are detected by performing a selected set of measurements on the system. These measurements are selected so that they do not damage the logical information, e.g.~measuring whether the state is still in the logical subspace. Outcomes of measurements contain the classical information, i.e.~error syndromes, that we need for correcting errors. Using these outcomes, we can work out the most likely errors and correct errors accordingly. With a better knowledge of the probability distribution of errors, we have a better chance to guess errors correctly, i.e.~error correction fails with a lower probability.

It is a common phenomenon that the strength of noise is determined by some time-dependent factors. For example, the amplitude of mechanical noise is usually time-dependent, e.g.~seismic noise depends on human activities, which can affect many experimental systems. For superconducting qubits, fluctuations in the population of unpaired electrons can lead to large temporal variations in the decoherence rate~\cite{Gustavsson2016}. In ion traps, the electric-field noise can excite phonons, which reduce the gate fidelity~\cite{Bermudez2017}, so the gate fidelity decreases with time between cooling operations. Drifts in the magnetic field and laser frequency can also cause time-dependent decoherence rate and gate fidelity in ion traps.

In this paper, we propose a protocol for estimating time-dependent error rates of each type of errors during quantum error correction. This protocol is based on processing error correction data using a machine learning algorithm, i.e.~online Gaussian processes~\cite{Csato2002}. Conventional approaches for measuring the noise in a quantum information device are the quantum process tomography~\cite{Nielsen2010} and randomised benchmarking~\cite{Knill2008}, however, both of them cannot be implemented when quantum error correction is in processing. Quantum error correction itself can be used to assess the noise~\cite{Fowler2014, Combes2014, Fujiwara2014}. For example, rates of error syndromes depend on error rates. Error syndrome rates can be used to reduce errors by optimising control parameters~\cite{Kelly2016} and overcoming systematic phase shifts~\cite{Muller2016}. Error syndrome rates can also be used to characterise quantum dynamics of a subsystem of the code~\cite{Mohseni2007}. Our protocol can access all qubits without interrupting the implementation of the code during scalable fault-tolerant quantum computing.

We propose two methods for estimating error rates based on correction operations and syndrome patterns, respectively. Both methods are approximate. The inaccuracy of the syndrome-pattern method is $\mathcal{O}(\epsilon^2)$, where $\epsilon$ is the error rate. Numerical results show that performances of two methods are similar. Using data obtained in error corrections in the past, the online Gaussian processes algorithm can predict error rates in the future, which are used in future error corrections to reduce error correction failures, i.e.~logical errors. Taking the surface code~\cite{Dennis2002} as an example, we find that both methods can significantly reduce logical errors with a factor that increases with the code distance. We remark that both extendible quantum memory and scalable quantum computation usually require a big code distance~\cite{Fowler2012, Joe2016, Dennis2002}. To implement our protocol, the required classical computing resource is $\mathcal{O}(n)$, where $n$ is the number of qubits. The algorithm for estimating error rates can be parallelised and runs in time $\mathcal{O}(1)$ with respect to $n$. Therefore, our protocol is realistic for scalable quantum information devices.

\begin{figure}[tbp]
\centering
\includegraphics[width=1\linewidth]{\figpath 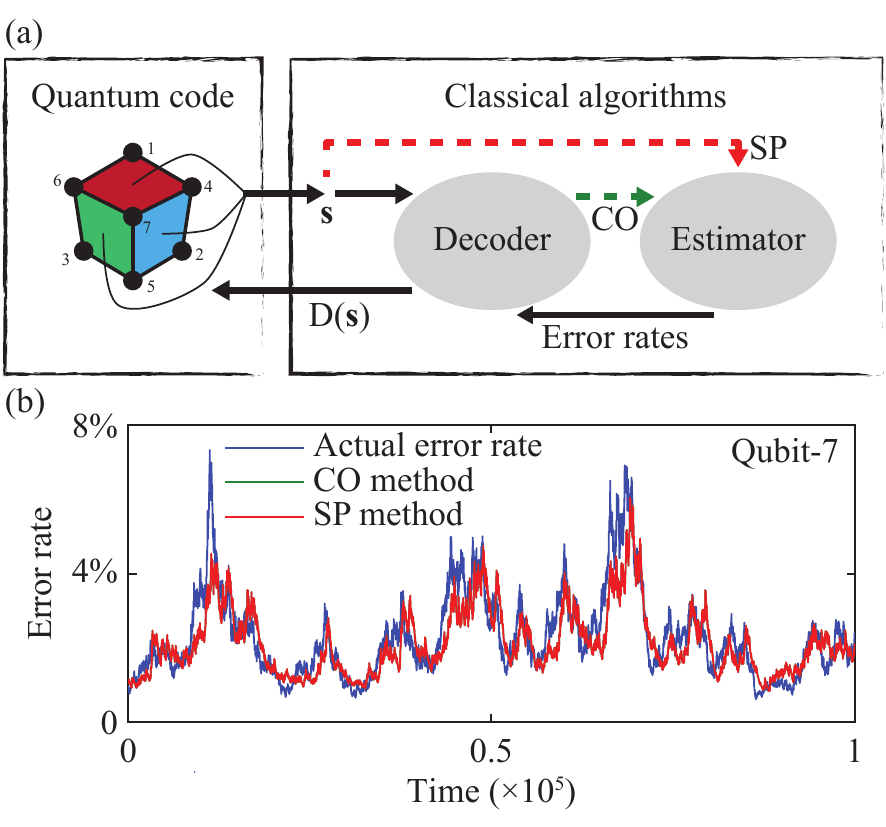}
\caption{
(a) Schematic diagram of quantum error correction enhanced by the real time error rate estimation. Taking the seven-qubit CSS code as an example, each circle denotes a qubit, and each face denotes two stabiliser operators $\SIG{\rm x}{}\SIG{\rm x}{}\SIG{\rm x}{}\SIG{\rm x}{}$ and $\SIG{\rm z}{}\SIG{\rm z}{}\SIG{\rm z}{}\SIG{\rm z}{}$ of four qubits on the face. The quantum computer sends stabiliser eigenvalues $\sss$ to the classical computer, and the classical computer returns correction operations $D(\sss)$. There are two methods for estimating error rates, which are the correction-operation (CO) method and the syndrome-pattern (SP) method. Arrows indicate the flow of the classical information.
(b) Time evolution of the error rate of qubit-7 in the seven-qubit CSS code. The decoder is the ideal decoder that uses error rates provided by the estimator to minimise the post-correction logical error probability. Data are generated numerically by taking ${\rm E}[\epsilon] = 0.02$ and $\sigma_\epsilon = 0.01$. The unit of time is one round of error correction. Error rates estimated using CO and SP methods coincide with each other. Error rates of other qubits are shown in Fig.~\ref{fig:Steane}.
}
\label{fig:figure1}
\end{figure}

\section{Quantum error correction}

We focus on stabiliser codes~(see Appendix). To encode $k$ logical qubits in $n$ physical qubits, usually the stabiliser of the code is an Abelian group generated by $n-k$ independent Hermitian Pauli operators (stabiliser generators). We use $\sss = (s_1, \ldots, s_{n-k})$, which is a string of numbers taking values $\pm 1$, to denote eigenvalues of these generators, and logical subspace is the subspace in which all eigenvalues are $+1$ (denoted by $\one$). Logical qubits are defined by logical Pauli operators $\{ \bar{\sigma} \}$. We consider Pauli errors $[\sigma]\rho = \sigma \rho \sigma^\dag$, where $\sigma$ is a Pauli operator. We use $\Sigma_\sss^{[\bar{\sigma}]}$ to denote the average of Pauli errors that flips stabiliser eigenvalues from $\one$ to $\sss$ and causes the logical Pauli error $[\bar{\sigma}]$ at the same time (see Appendix). Errors corresponding to the same set of $\sss$ and $[\bar{\sigma}]$ have the same effect on a logical state therefore do not have to be distinguished. Two errors occurring sequentially is equivalent to the error $\Sigma_{\sss_1}^{[\bar{\sigma}_1]}\Sigma_{\sss_2}^{[\bar{\sigma}_2]} = \Sigma_{\sss_1 \circ \sss_2}^{[\bar{\sigma}_1] [\bar{\sigma}_2]}$, where $\circ$ denotes the entrywise product, and $[\bar{\sigma}_1] [\bar{\sigma}_2] = [\bar{\sigma}_1 \bar{\sigma}_2]$.

Errors are detected by measuring $\sss$. If there is error $\Sigma_\sss^{[\bar{\sigma}]}$ on a state initially in the logical subspace, the measurement outcome is $\sss$. This outcome does not directly provide any information about the logical error $[\bar{\sigma}]$. To find out $[\bar{\sigma}]$, we need to implement the decoder in a classical computer. The decoder $D$ is a map from stabiliser eigenvalues $\sss$ to a logical error $[\bar{\sigma}]$. Given the outcome $\sss$, we decide that the error is $\Sigma_\sss^{D(\sss)}$, and an operation equivalent to $\Sigma_\sss^{D(\sss)}$ is performed to correct the error. The overall effect of both the error and the correction operation is $\Sigma_\sss^{D(\sss)} \Sigma_\sss^{[\bar{\sigma}]} = \Sigma_\one^{D(\sss) [\bar{\sigma}]}$, i.e.~the state is brought back to the logical subspace $\one$ with a logical error $D(\sss) [\bar{\sigma}]$. Therefore, error correction succeeds if $D(\sss) = [\bar{\sigma}]$ otherwise results in a logical error.

We consider the error model
\begin{eqnarray}
\tilde{\mathcal{N}} = \sum_{\sss, [\bar{\sigma}]} p_\sss^{[\bar{\sigma}]} \Sigma_\sss^{[\bar{\sigma}]},
\end{eqnarray}
which is a superoperator describing stochastic Pauli errors. Here, $p_\sss^{[\bar{\sigma}]}$ is the probability of error $\Sigma_\sss^{[\bar{\sigma}]}$. Due to this error model, error correction fails with the probability
\begin{eqnarray}
p_{\rm log} = 1 - \sum_\sss p_\sss^{D(\sss)},
\end{eqnarray}
which is the probability of logical errors after error correction. In order to minimise logical errors, we shall choose the correction operation according to the error with the maximum probability, i.e.
\begin{eqnarray}
D(\sss) = \mathrm{argmax}_{[\bar{\sigma}]} p_\sss^{[\bar{\sigma}]}.
\end{eqnarray}
Such an ideal decoder relies on the full knowledge of the error model. Without monitoring the noise in real time, we cannot optimise the performance of error correction if error probabilities are time-dependent. The knowledge of the error model is also important for optimising approximate decoders that cannot exactly minimise logical errors, e.g.~the surface code decoder using the minimum-weight perfect matching algorithm~\cite{Dennis2002}.

The flow of the real-time error rate estimation in error correction is shown in Fig.~\ref{fig:figure1}(a). In each round of error correction, the quantum computer measures stabiliser eigenvalues and sends outcomes to the classical computer. The classical computer runs two programs: the decoder and the error rate estimator. Using stabiliser eigenvalues, the decoder returns correction operations to the quantum computer. In the correction-operation method, the correction operation is decomposed by the decoder (as we will show later) and sent to the error rate estimator. In the syndrome-pattern method, stabiliser eigenvalues are directly sent to the estimator. Based on these data, the estimator provides real time estimations of error rates. Instead of optimising control parameters as in Ref.~\cite{Kelly2016}, we use these error rates to assist the decoder to increase the success probability of error correction.

\begin{figure}[tbp]
\centering
\includegraphics[width=1\linewidth]{\figpath 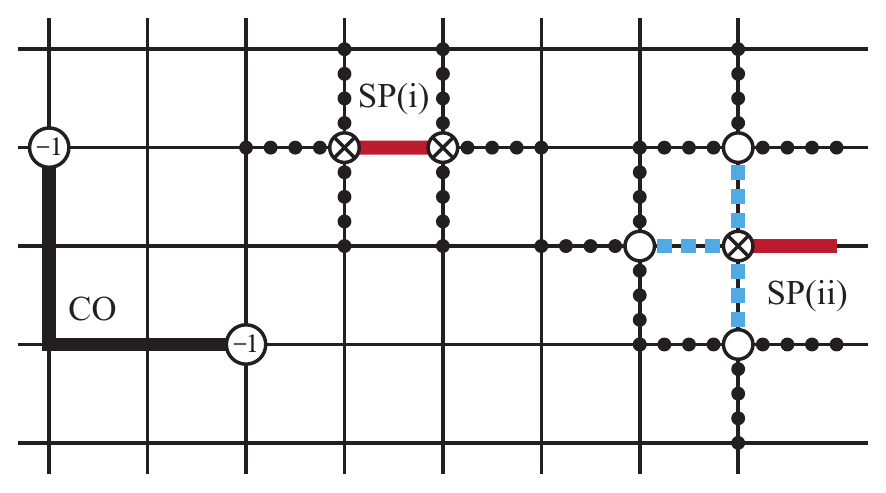}
\caption{
Two methods for observing error events in the surface code. Each vertex represents a stabiliser operator $\SIG{\rm x}{}\SIG{\rm x}{}\SIG{\rm x}{}\SIG{\rm x}{}$, and each edge represents a qubit or a phase-flip error on the qubit. In the correction-operation (CO) method, the decoder finds out correction operations on qubits (marked by the black bold line) that connecting stabiliser operators taking eigenvalue $-1$ (error syndromes, marked by circles with $-1$). These correction operations indicate errors on corresponding qubits. In the syndrome-pattern (SP) method, errors are observed using the pattern of error syndromes. The error $\Sigma_\sss^{[\bar{\sigma}]}$ is marked by the red bold line. The corresponding syndrome pattern $S_\sss$ contains these stabiliser operators marked by circles with crosses. $\mathcal{E}_\sss$ includes errors that can cause the same syndrome pattern, which are marked by the red bold line and blue dashed lines. $\tilde{S}_\sss$ includes stabiliser operators marked by circles with crosses and empty circles. $\tilde{\mathcal{E}}_\sss$ contains all these errors marked by the red bold line, blue dashed lines and also black dotted lines. In case-i, only one error can cause the syndrome pattern $S_\sss$, therefore $\mathcal{E}_\sss$ only contains one error, and $\tilde{S}_\sss = S_\sss$. In case-ii, the error $\Sigma_\sss^{[\bar{\sigma}]}$ is on a rough-boundary qubit, and these errors marked by blue dashed lines can case the same syndrome pattern as $\Sigma_\sss^{[\bar{\sigma}]}$. The pattern $S_\sss$ occurs faithfully if circles with crosses take the value $-1$ but empty circles take the value $+1$. To decide whether there is an error on the qubit marked by the red bold line, we check all stabiliser operators marked by circles. If and only if empty-circle stabiliser operators are +1 and crossed-circle stabiliser operators are -1 (syndromes), we decide that there is an error on the qubit. Using this strategy, we can make sure that, the syndrome pattern is either caused by the single-qubit error on the red bold line or multi-qubit errors.
}
\label{fig:figure2}
\end{figure}

\section{Indicators of error events}
\label{sec:methods}

Correction operations indicate error events. The stabiliser measurement outcome $\sss$ occurs with the probability $p_\sss = \sum_{[\bar{\sigma}]} p_\sss^{[\bar{\sigma}]}$, therefore the correction operation $\Sigma_\sss^{D(\sss)}$ is performed with the probability $p_\sss$. The probability distribution of correction operations can be described by
\begin{eqnarray}
\tilde{\mathcal{N}}_{\rm C} = \sum_{\sss, [\bar{\sigma}]} q_\sss^{[\bar{\sigma}]} \Sigma_\sss^{[\bar{\sigma}]},
\end{eqnarray}
where $q_\sss^{[\bar{\sigma}]} = \delta_{D(\sss), [\bar{\sigma}]} p_\sss$. The difference between $\tilde{\mathcal{N}}_{\rm C}$ and $\tilde{\mathcal{N}}$ (the probability distribution of errors) is limited by the post-correction logical error probability $p_{\rm log}$, i.e.~the distance
\begin{eqnarray}
\frac{1}{2} \sum_{\sss, [\bar{\sigma}]} \abs{q_\sss^{[\bar{\sigma}]} - p_\sss^{[\bar{\sigma}]}} = p_{\rm log}.
\end{eqnarray}
When error correction works properly, $p_{\rm log}$ is small. Therefore, two distributions are similar, and correction operations can approximately indicate error events, i.e.~when the correction operation is $\Sigma_\sss^{D(\sss)}$, it is likely that there is an error $\Sigma_\sss^{D(\sss)}$ on the state. However, there are $2^{n+k}$ terms in $\tilde{\mathcal{N}}$, so such a method is only practical when the qubit number $n$ is small, e.g.~short-distance repetition codes or the smallest surface code considered in Ref.~\cite{Fowler2014, Kelly2016}. In the following of this section, we develop two methods that are applicable to codes with many qubits, therefore these methods can be used in scalable fault-tolerant quantum computing.

The error model can be expressed in the product form
\begin{eqnarray}
\tilde{\mathcal{N}} = \prod_{\Sigma_\sss^{[\bar{\sigma}]} \in \mathcal{E}}
[ (1-\epsilon_\sss^{[\bar{\sigma}]})[\openone] + \epsilon_\sss^{[\bar{\sigma}]} \Sigma_\sss^{[\bar{\sigma}]}],
\end{eqnarray}
where $\openone$ is the identity operator, and $\{ \epsilon_\sss^{[\bar{\sigma}]} \}$ are error rates. Here, $\mathcal{E}$ is a subset of errors. When the noise is caused by $N$ independent sources, i.e.~$\tilde{\mathcal{N}} = \tilde{\mathcal{N}}_1 \cdots \tilde{\mathcal{N}}_N$, and each independent noise $\tilde{\mathcal{N}}_l$ only affects at most $M$ qubits, the overall noise $\tilde{\mathcal{N}}$ is determined by at most $4^M N$ error rates, i.e.~$\abs{\mathcal{E}} \leq 4^M N$. If these independent noises are all localised, e.g.~they are the decoherence of each individual qubit, we have $N = \mathcal{O}(n)$ and $M = \mathcal{O}(1)$ with respect to $n$. Error rates are real when the fidelity of each independent noise is greater than $\frac{1}{2}$. Although $\tilde{\mathcal{N}}$ is a completely positive map, $\epsilon_\sss^{[\bar{\sigma}]}$ may be negative. In the following, we assume that all $\epsilon_\sss^{[\bar{\sigma}]}$ are positive for simplification, and usually the magnitude of a negative error rate is small ($\sim \mathcal{O}(\epsilon^2)$). See Appendix for detailed discussions about the product form of the error model.

Usually, there are not two errors in $\mathcal{E}$ with the same subscript $\sss$. If $\Sigma_\sss^{[\bar{\sigma}_1]}, \Sigma_\sss^{[\bar{\sigma}_2]} \in \mathcal{E}$, the post-correction logical error probability $p_{\rm log} \geq \min\{ \epsilon_\sss^{[\bar{\sigma}_1]}, \epsilon_\sss^{[\bar{\sigma}_2]} \}$ (see Appendix). In this case $p_{\rm log}$ is not even lower than the rate of an independent error, i.e.~either the code is not proper, or only one of them (we suppose it is $\Sigma_\sss^{[\bar{\sigma}_1]}$) is significant and rates of others ($\Sigma_\sss^{[\bar{\sigma}_2]}$) are much lower. For the second case, we need to know which error is significant, so that errors with the same $\sss$ are not equally weighted by the decoder, otherwise $p_{\rm log}$ is on the level of the significant error rate ($\epsilon_\sss^{[\bar{\sigma}_1]}$) rather than insignificant error rates ($\epsilon_\sss^{[\bar{\sigma}_2]}$). Because rates of insignificant errors are on the level of the logical error probability, we can neglect them when estimating error rates. In the following, we assume that each $\sss$ corresponds to at most one unique error $\Sigma_\sss^{[\bar{\sigma}]}$ in $\mathcal{E}$.

The syndrome pattern of $\sss$ indicates the error $\Sigma_\sss^{[\bar{\sigma}]}$ with an inaccuracy $\mathcal{O}(\epsilon^2)$ (see Fig.~\ref{fig:figure2}). Without any error, all entries of $\sss$ (stabiliser operators) take the value $+1$. Error syndromes are entries taking the value $-1$, which indicate the presence of errors. Syndrome pattern $S_\sss$ is the set of all syndromes in $\sss$ (see Appendix for details). If $S_\sss \subseteq S_{\sss'}$, a syndrome in $\sss$ is also a syndrome in $\sss'$, and we say that the pattern $S_\sss$ occurs in $\sss'$. The set of errors $\mathcal{E}_\sss = \{ \Sigma_{\sss'}^{[\bar{\sigma}']} \in \mathcal{E} \mid S_\sss \subseteq S_{\sss'} \}$ includes errors in $\mathcal{E}$ that can cause the syndrome pattern $S_\sss$. We say that the pattern $S_\sss$ occurs {\it faithfully} in the stabiliser measurement outcome $\sss_{\rm o}$ if only $S_\sss$ occurs in $\sss_{\rm o}$ but syndrome patterns of other errors in $\mathcal{E}_\sss$ do not (i.e.~$S_\sss \subseteq S_{\sss_{\rm o}}$ but $S_{\sss'} \nsubseteq S_{\sss_{\rm o}}$ if $\Sigma_{\sss'}^{[\bar{\sigma}']} \in \mathcal{E}_\sss - \{\Sigma_\sss^{[\bar{\sigma}]}\}$). If $S_\sss$ occurs faithfully and it is caused by one error, the error must be $\Sigma_{\sss}^{[\bar{\sigma}]}$ rather than any other errors in $\mathcal{E}_\sss$. Only a subset of errors determines whether a syndrome pattern occurs faithfully. The set of syndromes $\tilde{S}_\sss = \bigcup_{\Sigma_{\sss'}^{[\bar{\sigma}']} \in \mathcal{E}_\sss } S_{\sss'}$ includes all syndromes that can be caused by errors in $\mathcal{E}_\sss$. We only need to check these syndromes in $\tilde{S}_\sss$ to find out whether $S_\sss$ occurs faithfully. The set of errors $\tilde{\mathcal{E}}_\sss = \{ \Sigma_{\sss'}^{[\bar{\sigma}']} \in \mathcal{E} \mid S_{\sss'} \bigcap \tilde{S}_\sss \neq \emptyset \}$ includes errors in $\mathcal{E}$ that can cause syndromes in $\tilde{S}_\sss$. All other errors in $\mathcal{E}$ but out of $\tilde{\mathcal{E}}_\sss$ are irrelevant to the syndrome pattern $S_\sss$. $S_\sss$ occurs faithfully (in the stabiliser measurement outcome) with the probability $\epsilon_\sss = \epsilon_\sss^{[\bar{\sigma}]} + \delta_\sss$, where $\epsilon_\sss^{[\bar{\sigma}]}$ is the rate of the corresponding error,
\begin{eqnarray}
\abs{\delta_\sss} \leq (P_\sss - \epsilon_\sss^{[\bar{\sigma}]})\epsilon_\sss^{[\bar{\sigma}]} + P_\sss^2,
\end{eqnarray}
and $P_\sss = \sum_{\Sigma_{\sss'}^{[\bar{\sigma}']} \in \tilde{\mathcal{E}}_\sss} \epsilon_{\sss'}^{[\bar{\sigma}']}$ (see Appendix). Therefore, the syndrome pattern can approximately indicate the corresponding error when $P_\sss$ is small.

Now, we propose two methods for observing error events.

\subsection{Correction-operation method}

In each round of quantum error correction, the quantum computer measures stabiliser eigenvalues to obtain $\sss_{\rm o}$ [see Fig.~\ref{fig:figure1}(a)]. Outcomes $\sss_{\rm o}$ are sent to the decoder in the classical computer in order to find out the correction operation $\Sigma_{\sss_{\rm o}}^{D(\sss_{\rm o})}$. In some decoders, e.g.~the minimum-weight perfect matching decoder of the surface code~\cite{Dennis2002} (see Fig.~\ref{fig:figure2}), the correction operation is automatically decomposed as a product of errors in $\mathcal{E}$ in the form
\begin{eqnarray}
\Sigma_{\sss_{\rm o}}^{D(\sss_{\rm o})} = \prod_{\Sigma_\sss^{[\bar{\sigma}]} \in \mathcal{E}}
(\Sigma_\sss^{[\bar{\sigma}]}) ^{\frac{1}{2}(1+y_{\sss,x}^{[\bar{\sigma}]})}.
\end{eqnarray}
Here, $y_{\sss,x}^{[\bar{\sigma}]} = +1$ ($-1$) indicates that there is an (no) error $\Sigma_\sss^{[\bar{\sigma}]}$ at the $x$-th round error correction. We remark that some decoders are not compatible with the correction-operation method, e.g.~the maximum-likelihood decoder of the surface code~\cite{Bravyi2014}. In this method, $\{ y_{\sss,x}^{[\bar{\sigma}]} \}$ also indicates how to correct errors. Data of error events $\{ y_{\sss,x}^{[\bar{\sigma}]} \}$ are sent to the error rate estimator for further processing [see Fig.~\ref{fig:figure1}(a)].

Probabilities of the overall correction operation and the actual overall error has a difference smaller than the post-correction logical error probability $p_{\rm log}$, but the difference between the probability of the observed error event $p(y_{\sss,x}^{[\bar{\sigma}]}=+1)$ and the actual rate of the error $\epsilon_\sss^{[\bar{\sigma}]}$ is not bounded by $p_{\rm log}$. In our numerical results, we find that the inaccuracy of this method is similar to the syndrome-pattern method.

\subsection{Syndrome-pattern method}

In this method, stabiliser measurement outcomes $\sss_{\rm o}$ are directly sent to the error rate estimator [see Fig.~\ref{fig:figure1}(a)], and the estimator observes error events by itself. The estimator notes that there is an error $\Sigma_\sss^{[\bar{\sigma}]}$ at the $x$-th round error correction (i.e.~$y_{\sss,x}^{[\bar{\sigma}]} = +1$) if the corresponding syndrome pattern $S_\sss$ occurs faithfully in $\sss_{\rm o}$, otherwise $y_{\sss,x}^{[\bar{\sigma}]} = -1$. We remark that, in this case errors are not corrected according to $\{ y_{\sss,x}^{[\bar{\sigma}]} \}$.

In this method, the probability of the observed error event is $p(y_{\sss,x}^{[\bar{\sigma}]}=+1) = \epsilon_\sss$, and the difference from the actual error rate $\epsilon_\sss^{[\bar{\sigma}]}$ is $\delta_\sss \sim \mathcal{O}(P_\sss^2)$. We note that $P_\sss \sim \abs{\tilde{\mathcal{E}}_\sss}\epsilon$. In the worst case scenario, $\tilde{\mathcal{E}}_\sss = \mathcal{E}$. When the noise is caused by localised sources, $\abs{\mathcal{E}} = \mathcal{O}(n)$, so $P_\sss \sim \mathcal{O}(n\epsilon)$ in this case. For extendible codes, e.g.~the surface code or color codes~\cite{Bombin2006}, $n$ can be a large number. However, usually the stabiliser of an extendible code has a structure that makes $\tilde{\mathcal{E}}_\sss = \mathcal{O}(1)$ with respect to $n$. For example, for the surface code, if errors are caused by the dephasing of each individual qubit, $\abs{\tilde{\mathcal{E}}_\sss} \leq 13$ and it is independent of $n$ (see Fig.~\ref{fig:figure2}).

\section{Error rate estimation}

Error rates are estimated using the data of error events, i.e.~$\{ y_{\sss,x}^{[\bar{\sigma}]} \}$. If error rates are time-independent, we can directly estimate them using the estimator $\hat{\epsilon}_\sss^{[\bar{\sigma}]} = N^{-1}\sum_{x = 1}^N \frac{1}{2}(1+y_{\sss,x}^{[\bar{\sigma}]})$, where $N$ is the total number of error correction rounds in the data. In this way, the uncertainty of the error rate estimation due to the shot noise is $\sqrt{\tilde{\epsilon}_\sss^{[\bar{\sigma}]}(1-\tilde{\epsilon}_\sss^{[\bar{\sigma}]})/N}$, where $\tilde{\epsilon}_\sss^{[\bar{\sigma}]} = p(y_{\sss,x}^{[\bar{\sigma}]}=+1)$ is independent of $x$. We remark that $\{ y_{\sss,x}^{[\bar{\sigma}]} \}$ is not genuine data of error events, because both methods for observing error events are not exact. Therefore the difference between the estimated error rate $\hat{\epsilon}_\sss^{[\bar{\sigma}]}$ and the actual error rate $\epsilon_\sss^{[\bar{\sigma}]}$ also depends on the inaccuracy of the method for observing error events.

Each error rate $\epsilon_\sss^{[\bar{\sigma}]}$ is estimated independently. Therefore, in the following we neglect the subscript $\sss$ and superscript $[\bar{\sigma}]$.

When the error rate is time dependent, estimating the error rate using the data of error events $\{ y_x \}$ is a typical classification problem that can be solved using the Gaussian processes approach. In this paper, we choose to use the online Gaussian processes algorithm~\cite{Csato2002}. Online Gaussian processes can be applied to large data sets with a reduced cost of computing resources. Some other online Bayesian methods have been used in parameter estimation and quantum tomography~\cite{Granade2016}. The online Gaussuan processes algorithm may provide an alternative tool in these tasks, e.g. estimating non-Markov processes.

Following the Gaussian processes classification approach, we assume that the error rate $\epsilon(f)$ is determined by a parameter $f$. The posterior distribution of the parameter $f$ is approximated by a Gaussian distribution determined by the average $\mean{f_x}_t$ and covariance $K_t(x,x')$. Here, $f_x$ is the value of $f$ at the $x$-round error correction, $K_t(x,x')$ is the covariance matrix of variables $f_x$ and $f_{x'}$, and the subscript $t$ denotes that the posterior is obtained using data from the first to $t$-th rounds of error correction. We note that $\mean{\cdot}_t$ denotes the average over the Gaussian distribution given by $\mean{f_x}_t$ and $K_t(x,x')$. For any function of $f_x$, we have
\begin{eqnarray}
\mean{F(f_x)}_t = \int df_x \frac{1}{\sqrt{2\pi K_t(x,x)}} e^{-\frac{(f_x - \mean{f_x}_t)^2}{2K_t(x,x)}} F(f_x).
\end{eqnarray}
In the online Gaussuan processes algorithm, $\mean{f_x}_t$ and $K_t(x,x')$ are calculated by iterating equations~\cite{Csato2002}
\begin{eqnarray}
\mean{f_x}_t &=& \mean{f_x}_{t-1} + q^{(t)}K_{t-1}(x,t), \label{eq:ff} \\
K_t(x,x') &=& K_{t-1}(x,x') \notag \\
&&+ r^{(t)}K_{t-1}(x,t)K_{t-1}(t,x'), \label{eq:KK}
\end{eqnarray}
where
\begin{eqnarray}
q^{(t)} &=& \frac{\partial}{\partial \mean{f_t}_{t-1}}\ln \mean{p(y_t | f_t)}_{t-1}, \\
r^{(t)} &=& \frac{\partial^2}{\partial \mean{f_t}_{t-1}^2}\ln \mean{p(y_t | f_t)}_{t-1},
\end{eqnarray}
and
\begin{eqnarray}
p(y_t | f_t) = \frac{1-y_t}{2} + y_t\epsilon(f_t).
\end{eqnarray}
To start the iteration, we have to introduce the prior Gaussian distribution $\mean{f_x}_0$ and $K_0(x,x')$, which are respectively the average and covariance before any data is used to estimate the distribution of $f$. By iterating equations, one set of data (i.e.~$y_t$) is added to the estimation each time. Because $y_t$ is contentiously generated in error correction, the posterior distribution is updated contentiously.

We can assume that the evolution of the parameter $f$, i.e.~the error rate $\epsilon$, is a Markov process and choose the prior distribution according to the Ornstein-Uhlenbeck process, i.e.~$K_0(x,x') = \sigma_f^2e^{-\frac{1}{\xi}\abs{x-x'}}$. Then the posterior distribution for $x,x'\geq t$ can be written as
\begin{eqnarray}
\mean{f_x}_t &=& \mean{f_x}_0 + e^{-(x-t)/\xi} \delta f_t, \label{eq:f} \\
K_t(x,x') &=& K_0(x,x') + e^{-(x-t)/\xi}e^{-(x'-t)/\xi} \delta K_t, \label{eq:K}
\end{eqnarray}
where $\delta f_t$ and $\delta K_t$ are calculated using
\begin{eqnarray}
\delta f_t &=& e^{-\frac{1}{\xi}}\delta f_{t-1} + q^{(t)}(K_0(t,t) + e^{-\frac{2}{\xi}}\delta K_{t-1}), \label{eq:df} \\
\delta K_t &=& e^{-\frac{2}{\xi}}\delta K_{t-1} + r^{(t)}(K_0(t,t) +  e^{-\frac{2}{\xi}}\delta K_{t-1})^2, \label{eq:dK}
\end{eqnarray}
and $\delta f_0 = \delta K_0 = 0$. Using data from the first to $t$-th rounds, the estimated error rate at the $x$-th round is $\hat{\epsilon}_x = \mean{p(y_x=+1 | f_x)}_t$.

The prior is taken to represent the prior knowledge about which kind of functions $f_x$ we expect. There are three parameters determining the Ornstein-Uhlenbeck process, $\mean{f_x}_0$, $\sigma_f$ and $\xi$. We need to measure these parameters before implementing the estimation by other means. Alternatively, we can optimise these parameters, for example, by minimising the logical error probability. The online Gaussian processes algorithm can be applied when the prior is not Markov, but the cost of computing resources may increase.

The data size $N$ increases with the running time of quantum error correction, because new error-event data are introduced at each round of error correction. Using the online Gaussian processes algorithm to predict the evolution of $f$, usually the number of parameters that need to be computed and stored increases quadratically with the data size~\cite{Csato2002}. Because the correlation in the prior covariance decreases with time, data collected long time ago ($\abs{x-x'}\gg \xi$) can be deleted from the dataset. Therefore, we can freeze the size of effective data once it is large enough.

When the evolution of error rates is Markovian and the prior covariance is set according to the Ornstein-Uhlenbeck process, only two parameters $\delta f_t$ and $\delta K_t$ need to be computed and stored for each error rate $\epsilon_\sss^{[\bar{\sigma}]}$. Every time new data are introduced, these two parameters are updated according to Eqs.~(\ref{eq:df},\ref{eq:dK}). Given $\delta f_t$ and $\delta K_t$, Eqs.~(\ref{eq:f},\ref{eq:K}) are used to predict the evolution of the error rate.

To estimate the error rate $\epsilon_\sss^{[\bar{\sigma}]}$, the classical computer needs to detect error events $\{y_{\sss,x}^{[\bar{\sigma}]}\}$. For the correction-operation method, detecting the error event is performed by the decoder and does not require any additional computing. For the syndrome-pattern method, identifying the syndrome pattern takes time $\mathcal{O}(1)$ and only needs to access $\mathcal{O}(1)$ stabiliser measurement outcomes. Therefore, the time cost is $\mathcal{O}(1)$ for estimating the error rate and updating the estimation. Each error rate $\epsilon_\sss^{[\bar{\sigma}]}$ is estimated independently, so the computing can be parallelised, and the overall required resource increases with the number of qubits as $\mathcal{O}(n)$. Here, we have assumed that noises are localised.

The error rate calculated using the $t$-th round data is not used in the $t$-th round error correction, because we do not want to use stabiliser measurement outcomes twice in one round of error correction.

If error corrections are performed periodically without a break, only the $(t+1)$-th round error correction uses the error rate given by $\mean{f_x}_t$ and $K_t(x,x')$, because these two parameters will be updated after new data are introduced at the $(t+1)$-th round. We remark that when stabiliser measurement outcomes may be incorrect due to measurement errors and gate errors that can affect measurement outcomes, we may need to repeatedly measure stabiliser operators for several times to perform one round of error correction. In many protocols of stabiliser-code-based fault-tolerant quantum computation, error correction on a subset of qubits needs to stop temporarily in order to inject magic states~\cite{Li2015} for completing the universality of logical quantum gates~\cite{Bravyi2005}, in which case we need to predict the error rate in a further future rather than only the next round.

At the beginning of quantum error correction, the error event data may not be adequate to estimate error rates. Therefore, we can introduce a warming up stage, in which the machine is not actually used in any computing task but error correction is implemented only for the purpose of collecting error-event data. After the warming up stage, we encode information in the error correction code to actually execute the computing task. Because of the ability to predict error rates in the future, error correction cycles can be broken temporarily to implement the encoding.

\begin{figure}[tbp]
\centering
\includegraphics[width=1\linewidth]{\figpath 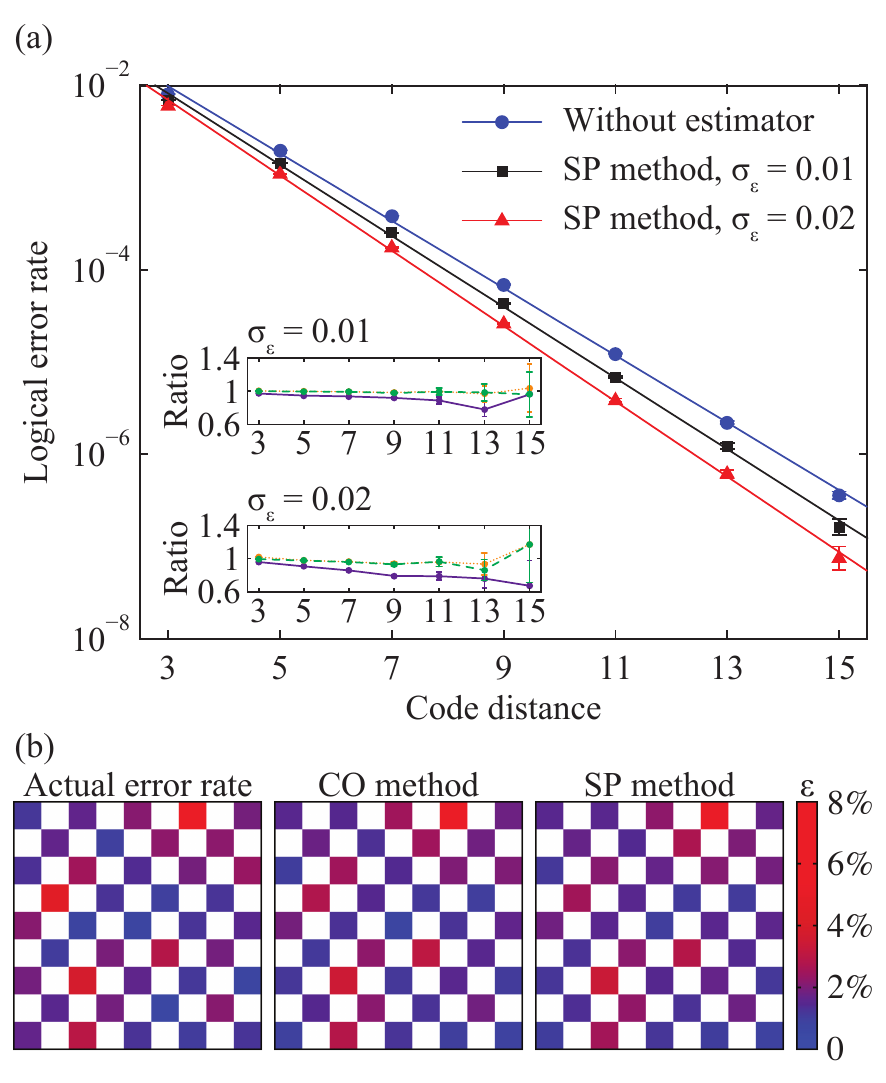}
\caption{
(a) Post-correction logical error probability $p_{\rm log}$ as a function of the code distance for the surface code. We take the expectation of the error rate ${\rm E}[\epsilon] = 0.02$, which is about one-fifth of the threshold of the surface code~\cite{Dennis2002}. The logical error probability decreases with the code distance. Data are calculated numerically using the Monte Carlo method, and lines are obtained by fitting data using Eq.~(\ref{eq:exp}). Without the error rate estimator, the logical error probability is independent of the variance of error rates $\sigma_\epsilon$. Using the syndrome-pattern (SP) method to estimate error rates, the logical error probability can be suppressed and decreases faster than without using the estimator. Insets: Ratios of logical error probabilities $p_{\rm log}'/p_{\rm log}$. Here, $p_{\rm log}'$ ($p_{\rm log}$) is the logical error probability obtained by using error rates estimated with the following method (SP method): For purple solid lines, we suppose that we can measure actual error rates; For the green dashed lines, error rates are estimated using the correction-operation method; For orange dotted lines, we suppose error-event data used by the estimator are genuine. The decoder is based on the minimum-weight perfect matching algorithm~\cite{Kolmogorov2009}, and weights are determined by estimated error rates~\cite{Dennis2002}. Error bars correspond to one standard deviation. (b) An example of actual error rates and error rates estimated using two methods for the surface code with the code distance $5$. Each square denotes a qubit, and the color represents the error rate.
}
\label{fig:figure3}
\end{figure}

\section{Numerical results}

To demonstrate the performance of our protocol, we numerically simulate quantum error corrections using the seven-qubit Calderbank-Shor-Steane (CSS) code~\cite{Steane1996} and the surface code, respectively. We consider the error model that errors are independent dephasing errors, i.e.
\begin{eqnarray}
\mathcal{N} = \prod_{j=1}^n [(1-\epsilon_j)[I] + \epsilon_j[\SIG{\rm z}{j}]].
\end{eqnarray}
Here, $\epsilon_j = \epsilon(f)$ is the probability of a dephasing error on qubit-$j$ occurring between two rounds of stabiliser measurements, $\epsilon(f) = (1-e^{-2e^f})/2$, and $f$ is time-dependent and different for each qubit. We take the dephasing error model as an example, but our methods can be applied to general error models as discussed in Sec.~\ref{sec:methods} and used to estimate the rate of correlated errors. Unlike dephasing caused by a slowly varying field that can be corrected using dynamical decoupling, the dephasing noise in this model can only be corrected using error correction. If the dephasing rate is $\gamma$ and the time between two rounds of stabiliser measurements is $\Delta t$, $\gamma\Delta t = e^f$. When $\gamma\Delta t \ll 1$, $\epsilon \simeq e^f$. Using this approximation, we calculate $q^{(t)}$ and $r^{(t)}$ by taking
\begin{eqnarray}
\mean{p(y_t | f_t)}_{t-1} &\simeq & \frac{1-y_t}{2} \notag \\
&&+ y_t \exp[\mean{f_t}_{t-1}+\frac{K_{t-1}(t,t)}{2}].
\end{eqnarray}
We assume that the evolution of the parameter $f$ is determined by an Ornstein-Uhlenbeck process, therefore using the Ornstein-Uhlenbeck process as the prior is exact in our examples. We choose $\mean{f_x}_0$ and $\sigma_f$ by taking the expectation of the error rate ${\rm E}[\epsilon] = 0.02$ and its standard deviation $\sigma_\epsilon = 0.01, 0.02$ (see Appendix), and we take the time scale of the relaxation in the Ornstein-Uhlenbeck process as $\xi = 5000$ rounds of error corrections.

In the seven-qubit CSS code, we can find that error rates estimated using two methods coincide with each other but are different from the actual error rate [see Fig.~\ref{fig:figure1}(b)]. Results of two methods are close because an average error rate much lower than the threshold is considered, in which case error syndromes are sparse, and the correction operation on a qubit is likely to be determined by only neighbouring error syndromes. The difference is due to the inaccuracy of the method for observing error events and also the inaccuracy of the online Gaussian processes algorithm~\cite{Csato2002}.

In the surface code, we demonstrate that the real time error rate estimation can reduce the post-correction logical error probability, and the improvement increases with the code distance [Fig.~\ref{fig:figure3}(a)]. When the code distance is larger, the probability $p_\sss^{[\bar{\sigma}]}$ of an error is determined by a product of more error rates, so choosing the optimal correction operation is more sensitive to the variance of each error rate. The logical error probability decreases exponentially with the code distance and can be approximately described using the formula
\begin{eqnarray}
p_{\rm log} = \exp(-\alpha d - \delta),
\label{eq:exp}
\end{eqnarray}
where $d$ is the code distance, and parameters $\alpha$ and $\delta$ are obtained by fitting numerical data in Fig.~\ref{fig:figure3}(a) (see Appendix). When $\alpha$ is larger, the logical error rate decreases faster with the code distance. We find that $\alpha = 0.8401$ without using the error rate estimator, which is increased to $\alpha = 0.8882$ ($0.9405$) by using the syndrome-pattern method when the $\sigma_\epsilon = 0.01$ ($0.02$). As shown in Fig.~\ref{fig:figure3}(b), error rates estimated using the correction-operation method and the syndrome-pattern method are similar. Differences between estimated error rates and actual error rates are plotted in Fig.~\ref{fig:surface}. Two methods have similar effects on the logical probability [see insets in Fig.~\ref{fig:figure3}(a)]. Data of error events obtained using both methods are not genuine. If we assume that data of error events are genuine (i.e.~$y_{\sss,x}^{[\bar{\sigma}]} = +1$ if and only if the corresponding error actually occurs), we find that the logical error probability is not further reduced, i.e.~for the purpose of reducing the logical error probability, both methods for observing error events have been optimised in this particular example. If we assume that actual error rates can be measured in some manner and used in error correction, the logical error probability can be further reduced, but the improvement is moderate.

\section{Conclusions}

By learning the time evolution of error rates in a quantum computer, quantum error correction can succeed with a higher probability. We have presented two methods for the real-time estimation of error rates, which work for any stabiliser code. Our methods can estimate error rates by processing the data generated in quantum error correction and do not require any additional qubits or operations in the quantum computer. Both methods are approximate but good enough to significantly reduce logical errors. Measuring error rates in error correction can also help us to understand the mechanism of errors and improve the fidelity of the quantum device. We have demonstrated both methods in the seven-qubit CSS code and the surface code. The seven-qubit CSS code is a candidate for demonstrating quantum error correction in the near term further, and the surface code is a promising code for the full-scale fault-tolerant quantum computation. We have only considered memory errors in our numerical simulations, but our methods can be applied to other error models, e.g.~each gate in the error correction circuit can cause errors. Our methods only require a small amount of classical computing resources for each qubit by using the online Gussian processes algorithm. In summary, our protocol is realistic and can significantly reduce logical errors in quantum error correction.

\begin{acknowledgments}
YL was supported by the EPSRC National Quantum Technology Hub in Networked Quantum Information Technologies. The authors would like to acknowledge the use of the University of Oxford Advanced Research Computing (ARC) facility in carrying out this work, http://dx.doi.org/10.5281/zenodo.22558.
\end{acknowledgments}

\appendix

\section{Stabiliser codes}

The Pauli group $G = \langle \SIG{\rm x}{1}, \SIG{\rm y}{1}, \SIG{\rm x}{1}, \ldots, \SIG{\rm x}{n}, \SIG{\rm y}{n}, \SIG{\rm x}{n} \rangle$ consists all Pauli operators of $n$ qubits together with multiplicative factors $\pm 1$ and $\pm i$, which has $4^{n+1}$ elements. Stabiliser $S$ is a subgroup of $G$. For a code encoding $k$ logical qubits in $n$ physical qubits, usually $S = \langle S_1, \ldots, S_{n-k} \rangle$ has $n-k$ independent generators. These generators are Hermitian and commute with each other. Each generator $S_i$ has two eigenvalues $\pm 1$. The $2^n$-dimensional Hilbert space is divided into $2^{n-k}$ subspaces according to stabiliser eigenvalues $\sss = (s_1, \ldots, s_{n-k})$. The projector to the subspace of $\sss$ reads $\pi_\sss = \prod_{i=1}^{n-k} \frac{1}{2}(I + s_i S_i)$, in which subspace the eigenvalue of $S_i$ is $s_i$. The dimension of each subspace is $2^k$. The logical subspace is the subspace that all eigenvalues are $s_i = +1$, and $k$ logical qubits are encoded in the logical subspace. We label the logical subspace with $\mathbf{1} = (1, \ldots, 1)$. These logical qubits are defined by logical Pauli operators, which are all elements of $G$. The Pauli group of $k$ logical qubits is $L = \langle \bar{\sigma}^{\rm x}_1, \bar{\sigma}^{\rm y}_1, \bar{\sigma}^{\rm z}_1, \ldots, \bar{\sigma}^{\rm x}_k, \bar{\sigma}^{\rm y}_k, \bar{\sigma}^{\rm z}_k \rangle$, which is a subgroup of $G$ and a subset of the centraliser of $S$, i.e.~elements of $L$ commute with all elements of $S$.

We consider Pauli errors in the form $[\sigma]\rho = \sigma\rho\sigma^\dag$, where $\sigma \in G$. These Pauli errors form an Abelian group $\mathcal{G} = \{[\sigma] \mid \sigma \in G\}$ with $4^n$ elements. We remark that $[\eta\sigma] = [\sigma]$, where $\eta = \pm 1, \pm i$. Similarly, $\mathcal{L} = \{[\bar{\sigma}] \mid \bar{\sigma} \in L\}$ is the subgroup of $4^k$ logical Pauli errors. $\mathcal{S} = \{[\sigma] \mid \sigma \in S\}$ is the subgroup of $2^{n-k}$ trivial errors: $[\sigma]\rho = \rho$ if the state $\rho$ is in the logical subspace (i.e.~$\pi_\one \rho = \rho$) and $[\sigma] \in\mathcal{S}$. There are $2^{n+k}$ cosets of $\mathcal{S}$ in the form $\mathcal{C}_\sss^{[\bar{\sigma}]} = [\sigma_\sss][\bar{\sigma}]\mathcal{S}$, where $\bar{\sigma} \in L$ is a logical operator, $\sigma_\sss \in G$ commutes with all logical operators, and $\sigma_\sss^\dag S_i \sigma_\sss = s_iS_i$. Because errors in $\mathcal{S}$ are trivial errors, errors in the same coset have the same effect on a logical state, i.e.~$[\sigma_1]\rho = [\sigma_2]\rho$ if the state $\rho$ is in the logical subspace and $[\sigma_1],[\sigma_2] \in \mathcal{C}_\sss^{[\bar{\sigma}]}$. We use $\Sigma_\sss^{[\bar{\sigma}]} = 2^{-n+k} \sum_{[\sigma] \in \mathcal{C}_\sss^{[\bar{\sigma}]}} [\sigma]$ to denote the average error. The average error $\Sigma_\sss^{[\bar{\sigma}]}$ and any element of $\mathcal{C}_\sss^{[\bar{\sigma}]}$ also have the same effect on a logical state, and $\{ \Sigma_\sss^{[\bar{\sigma}]} \} \cong \mathcal{G}/\mathcal{S}$.

\begin{figure*}[tbp]
\centering
\includegraphics[width=1\linewidth]{\figpath 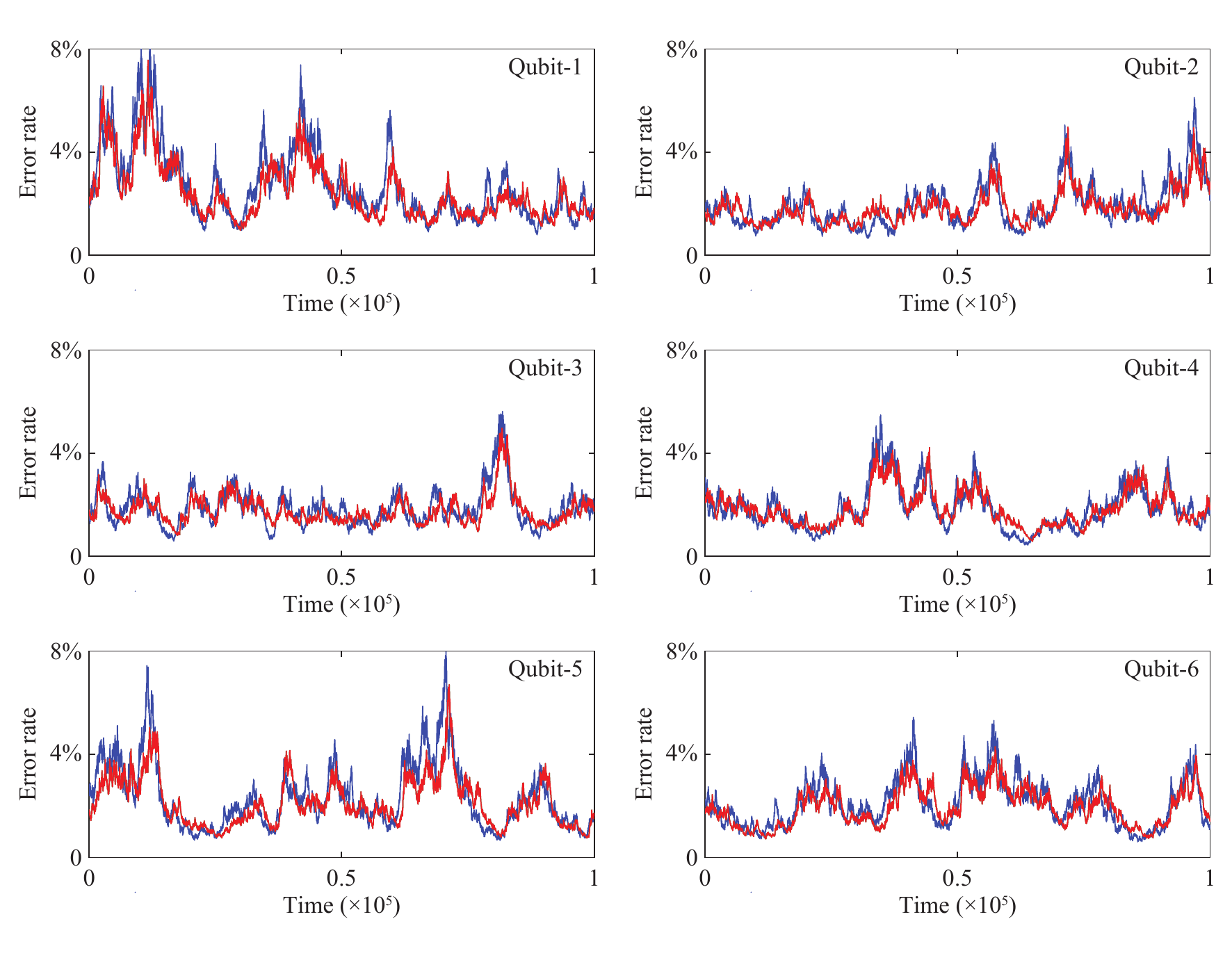}
\caption{
Time evolution of error rates of qubits 1 to 6 in the seven-qubit CSS code. Blue curves represent actual error rates, green curves represent error rates estimated using the correction-operation method, and red curves represent error rates estimated using the syndrome-pattern method. Error rates estimated using two methods coincide with each other. The unit of time is one round of error correction.
}
\label{fig:Steane}
\end{figure*}

\section{Error model}

The model of stochastic Pauli errors on $n$ qubits can be expressed as the noise superoperator $\mathcal{N} = \sum_{[\sigma] \in \mathcal{G}} p_{[\sigma]} [\sigma]$. Taking $p_\sss^{[\bar{\sigma}]} = \sum_{[\sigma] \in \mathcal{C}_\sss^{[\bar{\sigma}]}} p_{[\sigma]}$, $\mathcal{N}$ and $\tilde{\mathcal{N}} = \sum_{\sss, [\bar{\sigma}]} p_\sss^{[\bar{\sigma}]} \Sigma_\sss^{[\bar{\sigma}]}$ have the same effect on a logical state.

We consider the case that the noise includes $N$ independent sources of the noise and each independent noise only affects at most $M$ qubits. In this case $\mathcal{N} = \mathcal{N}_1 \cdots \mathcal{N}_N$, and each independent noise $\mathcal{N}_l = \sum_{[\sigma] \in \mathcal{G}_l} p_{l,[\sigma]} [\sigma]$, where $\mathcal{G}_l$ is the group of Pauli errors on these qubits affected by the $l$-th noise, and $\abs{\mathcal{G}_l} \leq 4^M$. We can reexpress an independent noise as (see the next section)
\begin{eqnarray}
\mathcal{N}_l = \prod_{[\sigma] \in \mathcal{G}_l} [ (1-\epsilon_{l,[\sigma]})[I] + \epsilon_{l,[\sigma]}[\sigma] ],
\end{eqnarray}
therefore the overall noise
\begin{eqnarray}
\mathcal{N} = \prod_{[\sigma] \in \cup_l \mathcal{G}_l} [ (1-\epsilon_{[\sigma]})[I] + \epsilon_{[\sigma]}[\sigma] ]
\end{eqnarray}
where $\abs{\bigcup_l \mathcal{G}_l} \leq 4^M N$. Using average errors, there is an equivalent error model in the product form,
\begin{eqnarray}
\tilde{\mathcal{N}} = \prod_{\Sigma_\sss^{[\bar{\sigma}]} \in \mathcal{E}} [ (1-\epsilon_\sss^{[\bar{\sigma}]})[I] + \epsilon_\sss^{[\bar{\sigma}]} \Sigma_\sss^{[\bar{\sigma}]}],
\end{eqnarray}
where $\mathcal{E} = \{ \Sigma_\sss^{[\bar{\sigma}]} \mid \mathcal{C}_\sss^{[\bar{\sigma}]} \bigcap (\bigcup_l \mathcal{G}_l) \neq \emptyset \}$ and $\epsilon_\sss^{[\bar{\sigma}]} \simeq \sum_{[\sigma]\in \mathcal{C}_\sss^{[\bar{\sigma}]} \cup (\cup_l \mathcal{G}_l)} \epsilon_{[\sigma]}$ when $\epsilon_{[\sigma]} \ll 1$.

We consider the case that $\Sigma_\sss^{[\bar{\sigma}_1]}, \Sigma_\sss^{[\bar{\sigma}_2]} \in \mathcal{E}$, where $[\bar{\sigma}_1] \neq [\bar{\sigma}_2]$. In this case, we can rewrite the error model as $\mathcal{N} = \mathcal{N}'_1 \mathcal{N}'_2 \mathcal{N}'$, where $ \mathcal{N}'_l = (1-\epsilon_\sss^{[\bar{\sigma}_l]})[I] + \epsilon_\sss^{[\bar{\sigma}_l]} \Sigma_\sss^{[\bar{\sigma}_l]}$, and $\mathcal{N}'$ describes the effect of all other errors in $\mathcal{E}$. If the error contributed by $\mathcal{N}'$ is $\Sigma_{\sss'}^{[\bar{\sigma}']}$, the overall error is one of $\Sigma_{\sss'}^{[\bar{\sigma}']}$, $\Sigma_{\sss' \circ \sss}^{[\bar{\sigma}'] [\bar{\sigma}_1]}$, $\Sigma_{\sss' \circ \sss}^{[\bar{\sigma}'] [\bar{\sigma}_2]}$ and $\Sigma_{\sss'}^{[\bar{\sigma}'] [\bar{\sigma}_1] [\bar{\sigma}_2]}$. When the stabiliser measurement outcome is $\sss'$, either $\Sigma_{\sss'}^{[\bar{\sigma}']}$ or $\Sigma_{\sss'}^{[\bar{\sigma}'] [\bar{\sigma}_1] [\bar{\sigma}_2]}$ or both of them can cause a post-correction logical error; similarly when the stabiliser measurement outcome is $\sss' \circ \sss$, either $\Sigma_{\sss' \circ \sss}^{[\bar{\sigma}'] [\bar{\sigma}_1]}$ or $\Sigma_{\sss' \circ \sss}^{[\bar{\sigma}'] [\bar{\sigma}_2]}$ or both of them can cause a post-correction logical error. Therefore, the post-correction logical error probability
\begin{eqnarray}
p_{\rm log} &\geq & \min\{ (1-\epsilon_\sss^{[\bar{\sigma}_1]})(1-\epsilon_\sss^{[\bar{\sigma}_2]}),  \epsilon_\sss^{[\bar{\sigma}_1]}\epsilon_\sss^{[\bar{\sigma}_2]} \} \notag \\
&&+ \min\{ (1-\epsilon_\sss^{[\bar{\sigma}_1]})\epsilon_\sss^{[\bar{\sigma}_2]}, (1-\epsilon_\sss^{[\bar{\sigma}_2]})\epsilon_\sss^{[\bar{\sigma}_1]} \}.
\end{eqnarray}
When $\epsilon_\sss^{[\bar{\sigma}_1]}, \epsilon_\sss^{[\bar{\sigma}_2]} \leq \frac{1}{2}$, we have $p_{\rm log} \geq \min\{ \epsilon_\sss^{[\bar{\sigma}_1]}, \epsilon_\sss^{[\bar{\sigma}_2]} \}$.

\begin{table}[tbp]
\begin{center}
\begin{tabular}{|c|c|c|}
\hline
$\sigma_\epsilon$ & $\mean{f_x}_0$ & $\sigma_f$ \\ \hline \hline
$0.01$ & $-4.0045$ & $0.4863$ \\ \hline
$0.02$ & $-4.2593$ & $0.8845$ \\ \hline
\end{tabular}
\end{center}
\caption{
Parameters $\mean{f_x}_0$ and $\sigma_f$. The expectation of the error rate ${\rm E}[\epsilon] = \mean{\epsilon(f)}_0 = 0.02$, and the standard deviation of the error rate $\sigma_\epsilon = \sqrt{\mean{\epsilon(f)^2}_0 - \mean{\epsilon(f)}_0^2}$.
}
\label{table}
\end{table}

\section{Product form of the error model}

\begin{figure}[tbp]
\centering
\includegraphics[width=1\linewidth]{\figpath 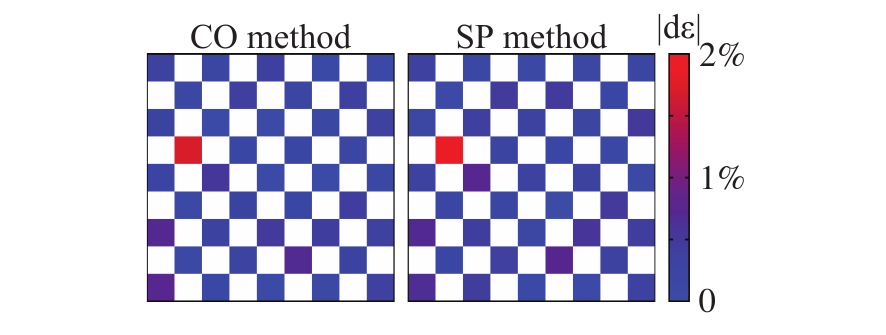}
\caption{
Differences between actual error rates $\epsilon$ and estimated error rates $\hat{\epsilon}$. The color of each square denotes the difference $\abs{d\epsilon}$ for a qubit of the surface code, where $d\epsilon = \epsilon - \hat{\epsilon}$.
}
\label{fig:surface}
\end{figure}

We consider the Pauli-error group $\mathcal{G}$ of $n$ qubits, and the conclusion can be applied to any subgroup, i.e.~$\mathcal{G}_l$ in the previous section. The group $\mathcal{G} = \langle \mathcal{P} \rangle$ has $2n$ independent generators, and $\mathcal{P} = \{ [\SIG{\rm x}{j}], [\SIG{\rm z}{j}] \mid j = 1,\ldots,n \}$. Each Pauli error $[\sigma] = \prod_{[\tau] \in \mathcal{P}} [\rho]^{b_{[\tau]}}$ has a unique expression using these generators, where $\mathbf{b}([\sigma]) = (b_{[\SIG{\rm x}{1}]}, b_{[\SIG{\rm z}{1}]}, \ldots, b_{[\SIG{\rm x}{n}]}, b_{[\SIG{\rm z}{n}]})$ is a string of $2n$ binary numbers (i.e.~$b_{[\tau]} = 0,1$).

We introduce superoperators
\begin{eqnarray}
\mathcal{A}_\mathbf{u} = \prod_{[\tau] \in \mathcal{P}} \frac{1}{2}([I]+i^{2u_{[\tau]}}[\tau]),
\end{eqnarray}
where $\mathbf{u} = (u_{[\SIG{\rm x}{1}]}, u_{[\SIG{\rm z}{1}]}, \ldots, u_{[\SIG{\rm x}{n}]}, u_{[\SIG{\rm z}{n}]})$ is also a string of $2n$ binary numbers. We can find that $\mathcal{A}_\mathbf{u} \mathcal{A}_{\mathbf{u}'} = \delta_{\mathbf{u},\mathbf{u}'} \mathcal{A}_\mathbf{u}$, and these $4^n$ superoperators $\{ \mathcal{A}_\mathbf{u} \}$ are linearly independent. Expanding $\mathcal{A}_\mathbf{u}$ as Pauli errors, we have
\begin{eqnarray}
\mathcal{A}_\mathbf{u} = 4^{-n} \sum_{[\sigma] \in \mathcal{G}}  i^{2\mathbf{u}\cdot \mathbf{b}([\sigma])} [\sigma],
\end{eqnarray}
where $\mathbf{u}\cdot \mathbf{b} = \sum_{[\tau] \in \mathcal{P}} u_{[\tau]} b_{[\tau]}$.

The noise $\mathcal{N} = \sum_{[\sigma] \in \mathcal{G}} p_{[\sigma]} [\sigma]$. Because of the linear independence, the noise can always be rewritten as
\begin{eqnarray}
\mathcal{N} = [I] + \sum_\mathbf{u} (\alpha_\mathbf{u}-1) \mathcal{A}_\mathbf{u},
\end{eqnarray}
where the summation covers all $\mathbf{u}$, and $\alpha_\mathbf{u} = \sum_{[\sigma] \in \mathcal{G}} i^{2\mathbf{u} \cdot \mathbf{b}([\sigma])}p_{[\sigma]}$. The coefficient $\abs{\alpha_\mathbf{u}} \leq 1$. Because the noise is trace-preserving, i.e.~$\sum_{[\sigma] \in \mathcal{G}} p_{[\sigma]} = 1$, we have $\alpha_\mathbf{u} \geq 0$ when $p_{[I]}\geq 1/2$.

Now, we are ready to reexpress the noise in the product form. Because $\mathcal{A}_\mathbf{u} \mathcal{A}_{\mathbf{u}'} = 0$ if $\mathbf{u} \neq \mathbf{u}'$, we can rewrite the noise as
\begin{eqnarray}
\mathcal{N} = \prod_\mathbf{u} [[I] + (\alpha_\mathbf{u}-1) \mathcal{A}_\mathbf{u}].
\end{eqnarray}
Because $\mathcal{A}_\mathbf{u}^2 = \mathcal{A}_\mathbf{u}$, we have $[I] + (\alpha_\mathbf{u}-1) \mathcal{A}_\mathbf{u} = e^{\mathcal{A}_\mathbf{u}\log\alpha_\mathbf{u}}$. Therefore $\mathcal{N} = e^{\sum_\mathbf{u} \mathcal{A}_\mathbf{u}\log\alpha_\mathbf{u}}$. Expanding $\{ \mathcal{A}_\mathbf{u} \}$ as Pauli errors, we have $\sum_\mathbf{u} \mathcal{A}_\mathbf{u}\log\alpha_\mathbf{u} = \sum_{[\sigma] \in \mathcal{G}} \beta_{[\sigma]} [\sigma]$, where $\beta_{[\sigma]} = 4^{-n} \sum_\mathbf{u} i^{2\mathbf{u}\cdot \mathbf{b}([\sigma])} \log\alpha_\mathbf{u}$. Then, we have the product expression of the noise
\begin{eqnarray}
\mathcal{N} = \prod_{[\sigma] \in \mathcal{G}} e^{\beta_{[\sigma]} ([\sigma]-[I])},
\end{eqnarray}
where
\begin{eqnarray}
e^{\beta_{[\sigma]} ([\sigma]-[I])} = e^{-\beta_{[\sigma]}} ([I]\cosh\beta_{[\sigma]} + [\sigma]\sinh\beta_{[\sigma]}),
\end{eqnarray}
i.e.~the error rate $\epsilon_{[\sigma]} = e^{-\beta_{[\sigma]}}\sinh\beta_{[\sigma]}$. Here, we have used that $\sum_{[\sigma] \in \mathcal{G}} \beta_{[\sigma]} = 0$, which is due to the fact that the noise is trace-preserving.

Although $\mathcal{N}$ is a completely positive map, $\epsilon_{[\sigma]}$ may be negative or even complex numbers. Usually, the overall error probability of the noise $1-p_{[I]}  \ll 1$ (which implies $p_{[I]}\geq 1/2$), in which case $\alpha_\mathbf{u} \geq 0$ and $\epsilon_{[\sigma]}$ is always real. When an error rate $\epsilon_{[\sigma]}$ is negative, usually the magnitude of $\epsilon_{[\sigma]}$ is much smaller than other error rates. Because $p_{[\sigma]} \sim \epsilon_{[\sigma]} + \mathcal{O}(\epsilon^2)$ is always non-negative, we have $\abs{\epsilon_{[\sigma]}} \sim \mathcal{O}(\epsilon^2)$ when $\epsilon_{[\sigma]}$ is negative. When expanding the product form, the product of two errors may result in the error $[\sigma]$ in the summation form, which leads to the term $\mathcal{O}(\epsilon^2)$. For example, we consider the noise of a single qubit in the form $\mathcal{N} = (1-2p)[I] + p [\SIG{\rm x}{}]  + p [\SIG{\rm z}{}]$, in which the probability of the $[\SIG{\rm y}{}]$ error is zero. For such a single-qubit noise, the product form reads $\mathcal{N} = \mathcal{N}_{\rm x} \mathcal{N}_{\rm y} \mathcal{N}_{\rm z}$, where $\mathcal{N}_{\rm \omega} = (1-2\epsilon_{\rm \omega})[I] + \epsilon_{\rm \omega} [\SIG{\rm \omega}{}]$, $\epsilon_{\rm x} = \epsilon_{\rm z} = (1-\sqrt{1-4p})/2$ and $\epsilon_{\rm y} = [1-(1-2p)/\sqrt{1-4p}]/2$. One can find that the rate of the $[\SIG{\rm y}{}]$ error is negative, and $\epsilon_{\rm y} = - \epsilon_{\rm x}^2/(1-2\epsilon_{\rm x})$. Regardless of the sign of $\alpha_\mathbf{u}$, in general $\log\alpha_\mathbf{u} = \log\abs{\alpha_\mathbf{u}} + i\pi \theta(-\alpha_\mathbf{u})$, where $\theta(\cdot)$ is the Heaviside step function. Therefore, the product expression always exists, but $\epsilon_{[\sigma]}$ may be complex numbers. In this paper, we focus on the case that $\epsilon_{[\sigma]}$ is always positive.

% $$\alpha_{00} = 1$$
% $$\alpha_{01} = 1-2p$$
% $$\alpha_{10} = 1-2p$$
% $$\alpha_{11} = 1-4p$$
% $$\beta_{00} = 4^{-1}[2\log(1-2p) + \log(1-4p)]$$
% $$\beta_{01} = 4^{-1}[- \log(1-4p)]$$
% $$\beta_{10} = 4^{-1}[- \log(1-4p)]$$
% $$\beta_{11} = 4^{-1}[-2\log(1-2p) + \log(1-4p)]]$$
% $$a = 4^{-1}[- \log(1-4p)] = $$
% $$b = 4^{-1}[-2\log(1-2p) + \log(1-4p)]]$$
% $$e^a = (1-4p)^{-1/4}$$
% $$e^b = [(1-2p)^2/(1-4p)]^{-1/4}$$

\begin{table}[tbp]
\begin{center}
\begin{tabular}{|c|c|c|c|c|}
\hline
$\sigma_\epsilon$ & $\alpha$ & $\sigma_\alpha$ & $\delta$ & $\sigma_\delta$ \\ \hline \hline
$0$ & $0.8401$ & $0.0126$ & $2.1078$ & $0.1242$ \\ \hline
$0.01$ & $0.8882$ & $0.0116$ & $2.1458$ & $0.1142$ \\ \hline
$0.02$ & $0.9405$ & $0.0103$ & $2.1524$ & $0.1019$ \\ \hline
\end{tabular}
\end{center}
\caption{
Parameters $\alpha$ and $\delta$ and their standard deviations. $\sigma_\epsilon = 0$ corresponds to the case that the error rate estimator is not used.
}
\label{tableII}
\end{table}

\section{Inaccuracy of the syndrome-pattern method}

Each error $\Sigma_\sss^{[\bar{\sigma}]}$ maps to a set of stabiliser generators $S_\sss = \{S_k \mid s_k = -1\}$ flipped by the error. $S_\sss$ is the syndrome pattern of the error $\Sigma_\sss^{[\bar{\sigma}]}$. The error $\Sigma_\sss^{[\bar{\sigma}]}$ occurs with the rate $\epsilon_\sss^{[\bar{\sigma}]}$. The syndrome pattern $S_\sss$ occurs faithfully with the probability $\epsilon_\sss = \sum_{\sss_{\rm o}} g_{\rm f}(\sss,\sss_{\rm o}) p_{\sss_{\rm o}}$. Here,
\begin{eqnarray}
g_{\rm f}(\sss,\sss_{\rm o}) = g(\sss,\sss_{\rm o}) \prod_{\Sigma_{\sss'}^{[\bar{\sigma}']} \in \mathcal{E}_\sss - \{\Sigma_\sss^{[\bar{\sigma}]}\}} (1-g(\sss',\sss_{\rm o})),
\end{eqnarray}
$g(\sss,\sss_{\rm o}) = \delta_{\frac{\mathbf{1}-\sss}{2} \cdot \frac{\mathbf{1}-\sss_{\rm o}}{2} , \abs{\frac{\mathbf{1}-\sss}{2}}}$, and $\abs{\cdot}$ denotes the number of ones in a binary string. $g(\sss,\sss_{\rm o})$ takes the value $1$ (otherwise $0$) if $S_\sss$ occurs in $\sss_{\rm o}$; $g_{\rm f}(\sss,\sss_{\rm o})$ takes the value $1$ (otherwise $0$) if $S_\sss$ occurs faithfully in $\sss_{\rm o}$.

Only errors in $\tilde{\mathcal{E}}_\sss$ determine whether $S_\sss$ occurs faithfully. To expand the product form of the noise, we introduce a binary string $\mathbf{v}$, in which $v_{\sss'} = 1$ ($v_{\sss'} = 0$) denotes that the error $\Sigma_{\sss'}^{[\bar{\sigma}']}$ is switched on (switched off). Each $\sss'$ is unique in $\mathcal{E}$, so each element of $\mathbf{v}$ is only labelled by $\sss'$. Then, we can rewrite the expression of $\epsilon_\sss$ by expanding the product form of the noise, and $\epsilon_\sss = \sum_{\mathbf{v}} g_{\rm f}(\sss,\sss_{\rm o}(\mathbf{v})) p(\mathbf{v})$,
\begin{eqnarray}
p(\mathbf{v}) &=& \prod_{\Sigma_{\sss'}^{[\bar{\sigma}']} \in \tilde{\mathcal{E}}_\sss} (1-\epsilon_{\sss'}^{[\bar{\sigma}']})^{1-v_{\sss'}} (\epsilon_{\sss'}^{[\bar{\sigma}']})^{v_{\sss'}}, \\
\sss_{\rm o}(\mathbf{v}) &=& \prod_{\Sigma_{\sss'}^{[\bar{\sigma}']} \in \tilde{\mathcal{E}}_\sss} (\sss')^{v_{\sss'}}.
\end{eqnarray}
In the expression of $\sss_{\rm o}(\mathbf{v})$, the product is the entrywise product. In this expression of $\epsilon_\sss$, only one term is non-zero when $\abs{\mathbf{v}} \leq 1$ ($\abs{\mathbf{v}}$ is the number of errors), which corresponds to the error $\Sigma_\sss^{[\bar{\sigma}]}$. Therefore, we can further rewrite $\epsilon_\sss = \epsilon_\sss^{(1)} + \epsilon_\sss^{(2)}$, where
\begin{eqnarray}
\epsilon_\sss^{(1)} &=& p_0\epsilon_{\sss}^{[\bar{\sigma}]}/(1-\epsilon_{\sss}^{[\bar{\sigma}]}), \\
\epsilon_\sss^{(2)} &=& \sum_{\mathbf{v} \mid \abs{\mathbf{v}}\geq 2} g_{\rm f}(\sss,\sss_{\rm o}(\mathbf{v})) p(\mathbf{v})
\end{eqnarray}
and
\begin{eqnarray}
p_0 = \prod_{\Sigma_{\sss'}^{[\bar{\sigma}']} \in \tilde{\mathcal{E}}_\sss} (1-\epsilon_{\sss'}^{[\bar{\sigma}']}).
\end{eqnarray}
We have
\begin{eqnarray}
\epsilon_\sss^{(2)} \leq \sum_{\mathbf{v} \mid \abs{\mathbf{v}}\geq 2} p(\mathbf{v}) = 1-p_0-p_1,
\end{eqnarray}
where
\begin{eqnarray}
p_1 = \sum_{\Sigma_{\sss'}^{[\bar{\sigma}']} \in \tilde{\mathcal{E}}_\sss} p_0\epsilon_{\sss'}^{[\bar{\sigma}']}/(1-\epsilon_{\sss'}^{[\bar{\sigma}']}).
\end{eqnarray}

The difference between $\epsilon_\sss^{[\bar{\sigma}]}$ and $\epsilon_\sss$ is $\delta_\sss = \epsilon_\sss - \epsilon_\sss^{[\bar{\sigma}]} = \epsilon_\sss^{(1)} + \epsilon_\sss^{(2)} - \epsilon_\sss^{[\bar{\sigma}]}$. Because $1-(1-a)(1-b)\leq a+b$ if $a,b\geq 0$, we have $1-p_0 \leq P_\sss$, where $P_\sss = \sum_{\Sigma_{\sss'}^{[\bar{\sigma}']} \in \tilde{\mathcal{E}}_\sss} \epsilon_{\sss'}^{[\bar{\sigma}']}$. Similarly, $1-p_0/(1-\epsilon_\sss^{[\bar{\sigma}]}) \leq P_\sss - \epsilon_\sss^{[\bar{\sigma}]}$. We also have $p_1 \geq p_0 P_\sss$. Therefore,
\begin{eqnarray}
\epsilon_\sss^{[\bar{\sigma}]} - \epsilon_\sss^{(1)} &=& [1-p_0/(1-\epsilon_\sss^{[\bar{\sigma}]})]\epsilon_\sss^{[\bar{\sigma}]} \notag \\
&\leq & (P_\sss - \epsilon_\sss^{[\bar{\sigma}]})\epsilon_\sss^{[\bar{\sigma}]}
\end{eqnarray}
and
\begin{eqnarray}
\epsilon_\sss^{(2)} \leq P_\sss - (1-P_\sss) P_\sss = P_\sss^2.
\end{eqnarray}
Both $\epsilon_\sss^{[\bar{\sigma}]} - \epsilon_\sss^{(1)}$ and $\epsilon_\sss^{(2)}$ are positive, therefore $\abs{\delta_\sss} \leq (P_\sss - \epsilon_\sss^{[\bar{\sigma}]})\epsilon_\sss^{[\bar{\sigma}]} + P_\sss^2$.

\section{Numerical data}

Data of qubits 1 to 6 in the seven-qubit CSS code are given in Fig.~\ref{fig:Steane}. To obtain these numerical results, we have taken $\mean{f_x}_0$ and $\sigma_f$ as shown in Table~\ref{table}. Fitting parameters $\alpha$ and $\delta$ are given in Table~\ref{tableII}. Differences between actual error rates and error rates estimated using the correction-operation method and the syndrome-pattern method are plotted in Fig.~\ref{fig:surface}.

\end{document}